\documentclass[authoryear,5p]{elsarticle}
\usepackage{amssymb}
\usepackage{amsmath} % \boldsymbol
\usepackage{hyperref} % gives \url
\usepackage{graphics}
\usepackage{color}
\definecolor{codegreen}{rgb}{0,0.6,0}
\definecolor{deepgreen}{rgb}{0,0.5,0}
\definecolor{deepred}{rgb}{0.9,0,0}
\definecolor{gray}{rgb}{0.4,0.4,0.4}
\definecolor{lightgray}{rgb}{0.9,0.9,0.9}
\usepackage{listings} % gives lstlisting
\usepackage{lipsum}
\newcommand\blfootnote[1]{%
  \begingroup
  \renewcommand\thefootnote{}\footnote{#1}%
  \addtocounter{footnote}{-1}%
  \endgroup
}
\usepackage{textcomp} % gives \textcopyright

\lstdefinestyle{mystyle}{
basicstyle=\ttfamily,
emph={self,Model,MulensModel,SatelliteSkyCoord,MODULE_PATH,MulensData,Event},
emphstyle=\color{deepred},
stringstyle=\color{codegreen},
keywordstyle=\color{magenta},
showstringspaces=false,
commentstyle=\color{gray}
}
\usepackage{lineno}

\newcommand\MM{{\tt MulensModel}}
\journal{Astronomy and Computing}

\begin{document}

\begin{frontmatter}

%% Title, authors and addresses

%% use the tnoteref command within \title for footnotes;
%% use the tnotetext command for the associated footnote;
%% use the fnref command within \author or \address for footnotes;
%% use the fntext command for the associated footnote;
%% use the corref command within \author for corresponding author footnotes;
%% use the cortext command for the associated footnote;
%% use the ead command for the email address,
%% and the form \ead[url] for the home page:
%%
%% \title{Title\tnoteref{label1}}
%% \tnotetext[label1]{}
%% \author{Name\corref{cor1}\fnref{label2}}
%% \ead{email address}
%% \ead[url]{home page}
%% \fntext[label2]{}
%% \cortext[cor1]{}
%% \address{Address\fnref{label3}}
%% \fntext[label3]{}

\title{Modeling microlensing events with \MM}

%% use optional labels to link authors explicitly to addresses:
%% \author[label1,label2]{<author name>}
%% \address[label1]{<address>}
%% \address[label2]{<address>}

% \author[label1]{}
\author{Rados{\l}aw Poleski\fnref{label1}}
\author{Jennifer C. Yee\fnref{label2}}

\address[label1]{Department of Astronomy, Ohio State University, 140 W. 18th Avenue, Columbus, OH 43210, USA}
\address[label2]{Harvard-Smithsonian Center for Astrophysics, 60 Garden Street, Cambridge, MA 02138 USA}

\begin{abstract}
We introduce \MM, a software package for gravitational microlensing modeling. 
The package provides a framework for calculating microlensing model magnification curves 
and goodness-of-fit statistics for microlensing events with single and binary lenses 
as well as a variety of
higher-order effects:  extended sources with limb-darkening, 
annual microlensing parallax, satellite microlensing parallax, 
and binary lens orbital motion.  The software could also be used for 
analysis of the planned microlensing survey by
the NASA flag-ship \emph{WFIRST} satellite.  
\MM\, is available at \url{https://github.com/rpoleski/MulensModel/}.
\end{abstract}

\begin{keyword}
%% keywords here, in the form: keyword \sep keyword
Gravitational microlensing [Exoplanets] \sep 
Gravitational microlensing \sep
Space telescopes \sep
Astroinformatics 
%% MSC codes here, in the form: \MSC code \sep code
%% or \MSC[2008] code \sep code (2000 is the default)

%% Methods: data analysis
\end{keyword}

\end{frontmatter}

% \linenumbers

%% main text
\section{Introduction
} % ######################################################
%\label{}
% (C) Year. 
\blfootnote{\textcopyright\,2018. This manuscript version is made available 
under the CC-BY-NC-ND 4.0 license \url{http://creativecommons.org/licenses/by-nc-nd/4.0/}
}
Gravitational microlensing allows the detection of massive bodies (lenses)
that are aligned with background sources of light.  
Light rays from the source 
are deflected by the lens gravity and the observer can see the increase in the flux
received from the source. The light-deflecting properties 
depend on the lens mass, not its 
luminosity, hence, the microlensing technique can be used to study a variety
of intrinsically faint objects that are hard to study using other methods:
exoplanets \citep[e.g., ][]{calchinovati15a,suzuki16}, 
brown dwarfs \citep[e.g., ][]{gould09,jung15}, 
distant low-mass stars \citep[e.g., ][]{skowron09,bensby17}, and 
stellar remnants \citep[e.g., ][]{mao02a,wyrzykowski16}.  
Gravitational microlensing was recently reviewed by \citet{mao12}, 
\citet{gaudi12}, and 
\citet{rahvar15}\footnote{We also recommend \url{http://microlensing-source.org/}}.  

An important capability of the microlensing technique is its ability to
study exoplanets \citep{mao91,gould92}.  Thanks to the natural scales 
(sources in the Galactic bulge and low-mass lenses a few kpc away), 
microlensing is sensitive
to planets on wide orbits \citep{han05a,poleski14c} and is able to detect 
planets with very low masses \citep{beaulieu06a,shvartzvald17b,bond17,udalski18b}.  
Microlensing signals can be detected due to planets that are not bound 
to any star, also called free-floating planets \citep{sumi11,mroz17b,mroz18a}.  
The statistical census of the planets with short orbital periods 
has been completed by the \emph{Kepler} mission \citep{coughlin16}.  
However, much less is known about the statistics of
planets on orbits similar to the ones observed in the Solar System, 
i.e., where most of the planetary 
mass is beyond the snow line ($2.7~{\rm AU}$).  
The need for much more detailed understanding of planetary systems beyond
the snow lines was recognized by the 2010 National Academy of Sciences' 
decadal survey and a microlensing survey from a satellite in space was 
recommended as a top priority for a large mission.  
The survey will be conducted by 
the \emph{Wide Field Infrared Survey Telescope} \citep[\emph{WFIRST}; ][]{spergel15} satellite. 

Here, we introduce \MM\, -- a new software package to compute microlensing 
light curves and compare them to observations.  One purpose of this 
package is to create a modern code
for microlensing that is easier to maintain and understand. Such a
code will allow new users to begin fitting microlensing models
quickly. With \MM, a scientist can easily begin
microlensing research without writing a light-curve modeling code from
scratch.  

To understand the context of \MM\, in relation to existing code,
we begin by briefly reviewing those codes.
There are several existing public codes for calculating the
magnification of finite-source binary-lens events. These codes tend to focus on implementing a particular algorithm for calculating the
magnification for a source of finite extent.
\citet{dominik07} described the Adaptive Contouring method and released 
its implementation.  The mapmaking method was first presented by 
\citet{dong06}, then improved by \citet{dong09b} and \citet{poleski14a}, 
and finally publicly released by \citet{poleski14c}.
The version released by \citet{poleski14c} was capable of modeling only 
a limited set of binary-lens events and has not
been used in independent peer-reviewed publications, most probably
because of the significant modifications the user must make to the source code to adapt it to a particular event.
The advanced contour integration algorithm was described by \citet{bozza10} 
and its implementation (called {\tt VBBL}) was released in 2016. 
An important advantage of the \citet{bozza10} approach was the error control, 
which allows the
user to set a threshold for the precision of the magnification
calculation, thereby allowing efficient use of computer
resources.  Also of note is the public release of a novel approach to
finding the roots of a fifth-order complex polynomial (e.g., the
solutions to the binary lens equation) that was presented by 
\citet{skowron12}\footnote{\url{http://www.astrouw.edu.pl/~jskowron/cmplx_roots_sg/}}.  

In contrast, \MM\, does not focus on a particular method for
calculating the magnification. Rather, it implements several methods for this
calculation and allows the user to select which method to use and when
to apply it. In fact, it incorporates several of these existing
algorithms for calculating the magnification (see Section~\ref{sec:bin}).

Recently, \citet{bachelet17} released the {\tt pyLIMA} package. This
package is designed around several ``use cases'' including generating microlensing
models, fitting those models to data, and generating simulated
microlensing data (see Section 2 of that paper). The fitting routines
are intended to be used with only ``rudimentary'' knowledge.

As we will describe in Section~\ref{sec:goals}, the main goal of \MM\, 
is to generate a microlensing model and enable the user to utilize 
an arbitrary fitting routine to optimize the microlensing parameters. 
This goal has some overlap with the {\tt pyLIMA} use cases to enable 
the user to fit models to data. However, specifically allowing for 
an arbitrary fitting routine has resulted in distinct differences in 
the implementation approaches of these two packages. In particular, \MM\,
does not implement any fitting routines, but rather leaves the fitting
to the user. At the same time, it is simpler to implement an arbitrary 
$\chi^2$ fitting routine with \MM\, than with {\tt pyLIMA} 
(compare \MM\,\\ example\_02\_fitting.py to 
{\tt pyLIMA}\\ pyLIMA\_example\_4.py).

Further direct comparisons to pyLIMA are complicated by the ongoing 
development of both codes. The original release of pyLIMA as described 
in \citet{bachelet17} was for point source point lens events only. Since then, 
binary sources and binary lenses have been added to the main GitHub 
repository, but they have not been accompanied by new releases of the code. 
For the binary lens magnification calculation, 
{\tt pyLIMA} incorporates only {\tt VBBL}, while \MM\, includes both this and other methods as well.  
In Section~\ref{sec:perf}, we
discuss a few direct comparisons of the performance of the two
packages.

This paper accompanies the release of version 1.4.0 of \MM. \MM\, is written in Python 3 with an % XXX
object-oriented approach.  It incorporates several different methods
for calculating the magnification of a binary lens and provides a
framework for using multiple methods together. We begin by describing the goals of the code in Section~\ref{sec:goals}. In Section~\ref{sec:app}, 
we describe the basic implementation approach and the major classes.  
In Sections~\ref{sec:calc} and \ref{sec:bin}, we briefly review the microlensing 
calculations and underlying computation methods.  
Section~\ref{sec:perf} presents performance tests.
Additional code features are described in Section~\ref{sec:add}. 
In Section~\ref{sec:fut}, we present plans for 
the future.  \ref{sec:appA} provides the source codes used to prepare 
Figures~\ref{fig:1}, \ref{fig:2}, and \ref{fig:3}.  
In \ref{sec:appB} we discuss publicly available microlensing datasets.

\section{Philosophy and Goals}% ######################################################
\label{sec:goals}  

Now is a particularly opportune time for new researchers to join the
microlensing field.  First, there are now several publicly available
microlensing datasets (see \ref{sec:appB}) that may be modeled, 
e.g., using \MM.   Second, efficiently, robustly, and completely exploring 
the high-dimensional microlensing parameter space (10--30 dimensions in most 
cases) remains a problem of ongoing scientific interest with new degeneracies 
continuously being discovered 
\citep[c.f.][]{hwang18c_tmp,hwang18a,poleski18a}. We can expect that 
the high quality of \emph{WFIRST} photometry will 
uncover new unsolved modeling challenges. Thus, it is a good moment
to start preparing for \emph{WFIRST} data analysis by exploring new
approaches to solving microlensing light curves. 

The computational problem of fitting microlensing models to data is two-fold.  First, accurate calculation of the magnification curve (a
single model) can be computationally expensive by itself. Second, the
likelihood surface can be complex with multiple minima, thus
requiring the generation of many models. Therefore, effectively and robustly searching parameter space for the best model can be extremely slow.

The purpose of \MM\, is to enable the user to experiment with
different optimizations for fitting microlensing events. For generating a particular model, the user can choose among
different (built-in) methods for calculating the light curve
magnification or perhaps in the future add their own method.  For finding
the best-fit model for a particular dataset, the user should be able
to implement an arbitrary method for searching the likelihood
surface. These
different methods (both fitting and model calculation) will have
different effects on both speed and accuracy (usual one at the expense
of the other). It is up to the user to determine what is best for
their application.

As a result of this high-level purpose, the goal of \MM\, is to
be as transparent to the user as possible. This means that \MM\,
does not fit for the microlensing parameters. \MM\, will
perform a linear fit for the flux parameters required to scale a model
to a given dataset and calculate the $\chi^2$, but fitting for the
best microlensing parameters is the responsibility of the
user. Likewise, it is up to the user to specify which method(s) to use
to calculate the magnification curve. Also, in the interest of
transparency, \MM\, is written in Python, with the goals of
following good programming practices and having good documentation
(see also Section~\ref{sec:app}).

The choice not to incorporate fitting algorithms into \MM\, is deliberate. 
Microlensing model
fitting is subject to many kinds of discrete degeneracies. These include 
mathematical degeneracies such as the ecliptic parallax degeneracy 
\citep{smith03} and the degeneracy between the projected separation of 
the lens components and its reciprocal \citep{dominik99}. There may also be
observational \citep[due to gaps in the data; e.g.,][]{park14}, and
astrophysical \citep[e.g., binary source vs.~planet in the lens system;][]{gaudi98} degeneracies.  
Furthermore, even in the absence of discrete
degeneracies, the likelihood space is often complex and highly
correlated in the common parameterizations of microlensing models.

Thus, there are two reasons that \MM\, does not have any built-in fitting 
algorithm. First, it is extremely difficult to write an algorithm that is 
robust in the face of the many known degeneracies. Aside from known 
mathematical degeneracies, the origins of the degeneracies (and therefore 
the situations in which they are relevant) may not even be understood. 
The problem is worse for degeneracies that have yet to be discovered. While it 
is possible and straightforward to incorporate a fitting routine that will 
find {\it a model} that fits the data, it is much harder to guarantee that it 
finds {\it all models} that fit the data.
Thus, even if a basic fitting algorithm were implemented, run, and passed 
the relevant metrics for ``success'', it is possible (or even likely) that 
it would find only one of multiple degenerate solutions. This outcome would 
violate our goal of transparency because the user would perceive only that 
the fit was ``successful'' and that the model was a good representation of 
the data. 

For example, the current version of pyLIMA available on GitHub gives 
an example for fitting an event with the parallax effect 
(pyLIMA\_example\_5.ipynb, commit {\tt 7d2366a}). This fitting does not 
require any input from the user regarding the initial conditions for 
the microlensing parameters. The notebook can be run without any specialized 
knowledge. At the same time, it reports only one of the degenerate parallax 
solutions (arising from the ecliptic parallax degeneracy). The only way for 
the user to know that the second solution exists is to have prior knowledge 
of this degeneracy.

Because of the difficulty of creating a ``black box'' for fitting 
microlensing events, the user will always have to evaluate how well 
the fitting algorithm has explored parameter space and whether or not 
additional solutions exist. Therefore, we prefer to have the user specify how 
the fitting should be done to make it explicit that they are responsible for 
the robustness of the results. Of course, there is still no guarantee that 
the user will find all degenerate solutions. However, this approach is less 
dangerous than providing a fitting routine that fails in ways that are not 
transparent to the user (especially if it fails in ways that manifest as 
``success'').

The second reason for having the user specify the fitting algorithm rather 
than including one with \MM\, is that robust fitting of microlensing models to 
data is precisely the problem we hope \MM\, will be used to solve.
The Markov Chain Monte Carlo technique is commonly used to fit for the best 
microlensing parameters (and their uncertainties). However, there exist many 
other methods for parameter optimization and searching the likelihood space. 
A major goal of \MM\, is to create a package that can interface easily with 
an arbitrary optimization routine. 
The goal is to allow the user to experiment with and determine which algorithms are better or worse for exploring microlensing model parameter space.

Specific examples of
how \MM\, can be combined with optimization routines such as 
{\tt EMCEE} \citep{foremanmackey13} or 
{\tt MultiNest} \citep{feroz08,feroz09} 
are distributed together with the \MM\, code. 
The user may use these fitting examples as  ``black box'' 
fitting scripts that require just a few settings to be changed, but should keep in mind the issues discussed above.

Aside from this high level goal, we also have a goal that \MM\, should 
be useful for modeling \emph{WFIRST} photometry. This leads to the practical 
constraint that the \MM\, code should have numerical accuracy that is high 
enough to allow analysis of the \emph{WFIRST} data.

\section{Implementation approach} % ######################################################
\label{sec:app}  

\MM\, is developed and distributed via the GitHub platform:
\begin{center}
\url{https://github.com/rpoleski/MulensModel}
\end{center}
The version 1.4.0 can be accessed via: % XXX
\begin{center}
\url{https://github.com/rpoleski/MulensModel/releases/tag/v1.4.0} % XXX
\end{center}
The on-line documentation is available at:
\begin{center}
\url{https://rpoleski.github.io/MulensModel/}
\end{center}

\MM\, follows 
the PEP8\footnote{\url{https://www.python.org/dev/peps/pep-0008/}} 
coding standard.  
The numbering of code versions follows the Semantic
Versioning\footnote{\url{https://semver.org/}} scheme.  We use the Sphinx
environment\footnote{\url{http://www.sphinx-doc.org/}} to build the
documentation.  The Git\footnote{\url{https://git-scm.com/}} version control
system is used for development.  To load and run modules written 
in C we use portable Python module 
{\tt ctypes}\footnote{\url{https://docs.python.org/3.5/library/ctypes.html}}.  
\MM\, can be installed using python packaging tools.
To produce plots we call Python module {\tt Matplotlib} \citep{hunter07}.

\MM\, was designed based on a series of use cases, which were written as 
fully executable code. In general, a use case was written for a new feature 
before beginning the implementation of that feature. This method allowed us 
to determine the structure of the code (classes, methods, and various 
arguments) based on the desired API. These use cases are not included in 
the releases, since a number of them reflect features for future development 
and so do not work. However, they are available on the main Git repository. 

The reproducibility of the results is ensured by unit tests of which
we have more than 80 currently. 
Many of these unit tests were created from existing (but non-public) 
microlensing codes that are well-vetted and have been used for many published 
analyses.

The \MM\, user will interact primarily with following classes:
{\tt Model}, {\tt MulensData}, and {\tt Event}.  We describe them below.

The {\tt Model} class defines a microlensing model and handles the
calculation of the magnification curve.  A given model is specified by
a set of parameters stored in the {\tt ModelParameters} class; 
the parameters implemented in version 1.4.0 are given in Table~\ref{tab:var}.  % XXX
\MM\, follows the microlensing parameter
conventions defined by \citet[][Appendix A]{skowron11}, see also \citet{gould00b}.  
Specifically, $t_0$ and $u_0$ are defined relative to the
center of mass of the lens system, 
$t_{\rm E}$ is defined relative to the Einstein radius of
the total mass of the lens system, and $\alpha$ is measured
counter-clockwise from the binary lens axis to the source trajectory.  
The current version of the code uses the point source method as default for calculating the magnification.  
The {\tt Model} class allows the user to specify that a more accurate method 
for calculating the magnification be used for a particular time range. 
Multiple time ranges with different methods may be specified. 
The methods that can be selected are presented in the next two sections.

{\tt MulensData} is used to store a photometric dataset (i.e., epochs
and photometric measurements and their uncertainties).  
For each dataset, its properties
are specified independently: bandpass (used for source
limb-darkening), satellite ephemeris (if any), the format of
photometry (flux or magnitudes), and the mask of epochs
to be excluded from the calculations.  
{\tt MulensData} reads typical three-column ASCII files 
with photometric data. The user can also use various keywords to specify other file formats.  
Alternatively, the user may read the data into a python list or a {\tt numpy} array 
and pass these as arguments of {\tt MulensData}.

\begin{table*}
\begin{tabular}{l l l l}
Parameter & Name in &  Unit & Description \\
 & \MM &  & \\
\hline
$t_0$ & {\tt t\_0} & & The time of the closest approach between the source and the lens. \\ 
$u_0$ & {\tt u\_0} & & The impact parameter between the source and the lens center of mass. \\
$t_{\rm E}$ & {\tt t\_E} & d & The Einstein crossing time. \\
$t_{\rm eff}$ & {\tt t\_eff} & d & The effective timescale, $t_{\rm eff} \equiv u_0 t_{\rm E}$. \\
$\rho$ & {\tt rho} & & The radius of the source as a fraction of the Einstein ring. \\
$t_{\star}$ & {\tt t\_star} & d & The source self-crossing time, $t_\star \equiv \rho t_{\rm E}$. \\
$\pi_{{\rm E}, N}$ & {\tt pi\_E\_N} & & The North component of the microlensing parallax vector. \\
$\pi_{{\rm E}, E}$ & {\tt pi\_E\_E} & & The East component of the microlensing parallax vector. \\
$t_{0,{\rm par}}$ & {\tt t\_0\_par} & & The reference time for parameters in parallax models. \\
$s$ & {\tt s} & & The projected separation between the lens primary and its companion\\
 & & & as a fraction of the Einstein ring radius. \\
$q$ & {\tt q} & & The mass ratio between the lens companion and the lens \\
 & & & primary $q \equiv m_2/m_1$. \\
$\alpha$ & {\tt alpha} & deg. & The angle between the source trajectory and the binary axis. \\
$ds/dt$ & {\tt ds\_dt} & yr$^{-1}$ & The rate of change of the separation. \\
$d\alpha/dt$ & {\tt dalpha\_dt} & deg.~yr$^{-1}$ & The rate of change of $\alpha$. \\
$t_{0,{\rm kep}}$ & {\tt t\_0\_kep} & & The reference time for lens orbital motion calculations. \\
\end{tabular}
\caption{Names of microlensing parameters in \MM\, class {\tt ModelParameters}. 
\label{tab:var}
}
\end{table*}

The {\tt Event} class combines any number of instances of {\tt
MulensData} with an instance of {\tt Model}. The main method of the
{\tt Event} class is {\tt get\_chi2()}, which calculates the total
$\chi^2$ statistic for all datasets relative to the given {\tt Model}.
The observed flux $F(t)$ consists of the flux from the magnified source star
$F_{S}$ blended with unmagnified flux from other stars $F_{B}$.  
Since the point-spread-function of each dataset is unique and the microlensing 
events are mostly observed in highly crowded sky regions, 
the blend flux must be fitted for each dataset
independently.  We also fit the source flux to each dataset, 
since the photometric data are often 
uncalibrated, and hence the source flux is on an arbitrary photometric system.
The magnification $A(t)$ is calculated for each dataset, and
then 
\MM\, fits for $F_{S}$ and $F_{B}$ by finding 
least-squares solution of a linear matrix equation, defined by
\begin{equation} \label{eq:flux}
{\mathbf F}_{{\rm obs}} = {\mathbf A}_{\rm mod} F_{{\rm S}} + F_{{\rm B}},
\end{equation}
where ${\mathbf F}_{\rm obs}$ is the vector of observed fluxes and 
${\mathbf A}_{\rm mod}$ is the vector of model magnifications calculated 
for each data point.

\section{Point lenses} % ######################################################
\label{sec:calc}

In this section we review how the microlensing magnification is calculated 
for point-lens events (i.e., the lens is only a single mass and does not obscure the source). 
These calculations are implemented in the {\tt PointLens} class.

Because the point lens case is relatively tractable, we describe
the magnification calculations in some detail. As we will see in Section~\ref{sec:bin}, 
when these equations are extended to the two-body
case, the increased complexity results in significant computational
challenges for accurately calculating the magnification.

\subsection{Lens equation and Paczy\'nski curve} % ######################################################
\label{sec:calc1}

Magnification and the properties of the source images are derived from the lens equation that 
maps the source plane coordinates of a light ray to the image plane coordinates.  
Let $\beta$ be the angle between the unperturbed source position 
and the lens position, whereas $\theta$ is the image angular separation from 
the lens.  The lens equation is \citep[see, e.g.,][]{schneider92,gaudi12}:
\begin{equation} \label{eq:first}
\beta = \theta - \frac{4GM}{c^2}\frac{D_s-D_l}{D_s D_l}\frac{1}{\theta},
\end{equation}
where 
$G$ is the gravitational constant, 
$M$ is the lens mass, 
$c$ is the speed of light, 
$D_s$ is the source distance, and 
$D_l$ is the lens distance 
(both measured from the observer).  
The above equation can be written in a simpler form, once we introduce 
the Einstein ring radius $\theta_{\rm E}$, which characterizes the event as 
a whole:
$$
\theta_{\rm E} = \sqrt{\frac{4GM}{c^2}\frac{D_s-D_l}{D_s D_l}}.  
$$
Then Equation~\ref{eq:first} becomes:
$$
\beta = \theta - \frac{\theta_{\rm E}^2}{\theta}.
$$
We can further simplify the notation if we use $\theta_{\rm E}$ as 
a fundamental angular scale and substitute: $u = \beta/\theta_{\rm E}$ and 
$y = \theta/\theta_{\rm E}$:
$$
u = y - \frac{1}{y}
$$
The image magnification is $A = \frac{y}{u}\frac{dy}{du}$.  For a single lens 
there are two images and the total magnification is \citep{paczynski86}:
\begin{equation} \label{eq:Au}
A(u) = \frac{u^2+2}{u\sqrt{u^2+4}}.
\end{equation}
This equation is exact for a point source, and
as long as $u$ is much larger than the source size ($\rho$),  
a point source approximation
is sufficient to describe the magnification.  

\begin{figure}
\includegraphics[width=0.455\textwidth,trim={20 0 50 0},clip]{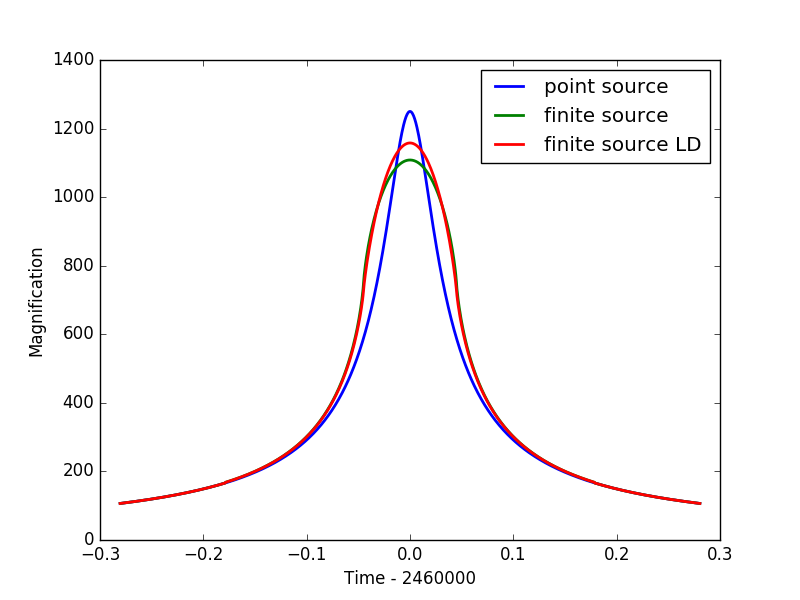}
\caption{Example point-lens magnification curves.  
Blue line shows the point-source approximation (Paczy\'nski curve) with 
high peak magnification.  Green curve represents 
the finite source magnification curve ($\rho = 0.0017$ and $u_0=0.0008$), 
while the red curve additionally includes 
the limb-darkening effect ($\Gamma = 0.4$).  
The source codes used to produce this figure and Figures~\ref{fig:2} and \ref{fig:3}
are provided in the \ref{sec:appA}.  The values of all the parameters 
are given in the first few lines of each source code. 
\label{fig:1}}
\end{figure}

\begin{figure}
\includegraphics[width=.95\columnwidth]{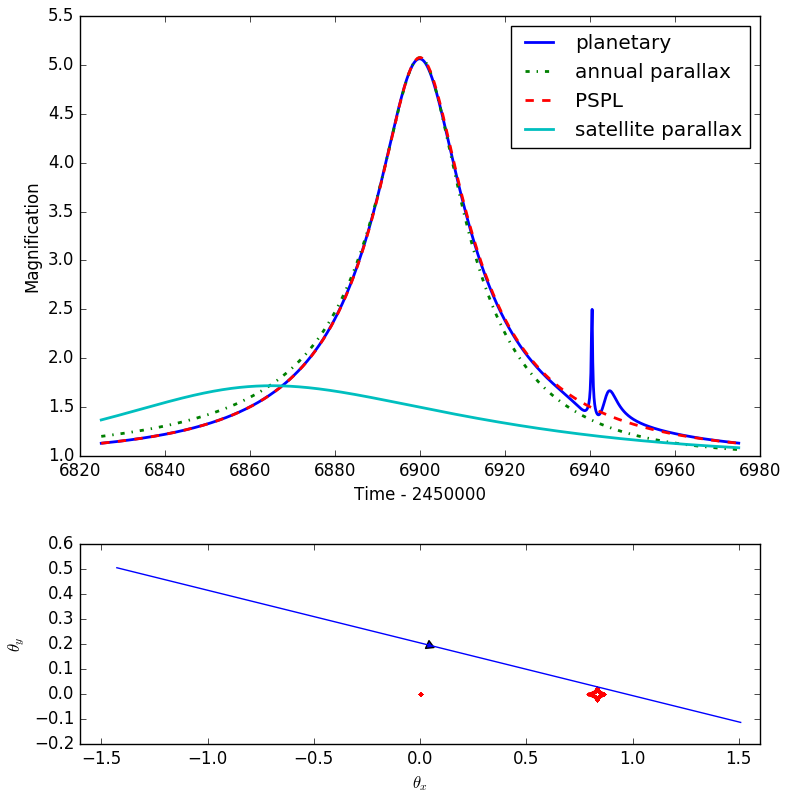}
\caption{Example magnification curves (top panel): 
dashed red -- point-source point-lens, 
solid blue -- binary lens with planetary mass ratio ($q=0.001$, $s=1.5$), 
dot-dashed green -- point-source point-lens curve affected by the annual parallax effect 
(relatively large value of $\pi_{\rm E} = 0.61$), and
solid cyan -- point-source point-lens curve as seen by 
the \emph{Spitzer} satellite (for the same parallax value). 
Note that the planetary model seen in the blue curve is significantly 
more complicated than a simple sum of two point-source point-lens models. 
The lower panel shows the source trajectory (blue) 
relative to caustics (red) for the planetary model from the top panel.
\label{fig:2}}
\end{figure}

When the relative lens-source motion can be approximated as 
rectilinear, then $u$ for an epoch $t$ is given by:
\begin{equation} \label{eq:u2}
u^2 = u^2_0 + \tau^2\newline
\end{equation}
\begin{equation} \label{eq:tau}
\tau \equiv \frac{t - t_0}{t_{\rm E}},
\end{equation}
where $t_0$ is the epoch of the minimum approach, 
$u_0$ is the impact parameter, 
and $t_{\rm E}$ is the Einstein timescale \citep{gould00b}.  
The Einstein timescale is defined as $t_{\rm E} = \theta_{\rm E}/\mu$, 
where $\mu$ is the lens-source relative proper motion.  
The system of Equations \ref{eq:Au}, \ref{eq:u2}, and \ref{eq:tau} 
are frequently called the Paczy\'nski curve.  
We present Paczy\'nski curves with high peak magnification in 
Figure~\ref{fig:1} (blue line) and low peak magnification in 
Figure~\ref{fig:2} (dashed red line).  

\subsection{Finite source} % ######################################################
\label{sec:calc2}

We see from Equation~\ref{eq:Au}, 
which is derived in the point-source point-lens limit, 
that the magnification would be infinite for $u=0$.  
In reality, the assumption of a point 
source is not valid, as every source has some finite, even if very small, 
angular size.  For a typical bulge source, the angular source radius 
($\theta_{\star}$) is a few ${\rm \mu arcsec}$, or about $0.001$ of 
$\theta_{\rm E}$ \citep[e.g.,][]{gaudi12}.  Hence, if $u$ is smaller than a few times 
$\rho \equiv \theta_{\star}/\theta_{\rm E}$, then different parts of the source disk 
are magnified by significantly different amounts and the magnification
differs from the point-source 
approximation \citep{wittmao94}.  \MM\, implements the finite source formalism 
by \citet{gould94b}, which was further extended to limb-darkened sources by 
\citet{yoo04b}.  These methods are accurate as long as $\rho \lesssim 0.1$.  
The functions needed for finite source calculations were precomputed and 
cubic spline interpolation is used to find requested value. We have checked 
that relative interpolation errors are below $10^{-4}$.  
Alternatively, the user may request direct integration of underlying functions. 
See Figure~\ref{fig:1} for example magnification curves with a finite source.  
The finite-source magnification curve is flattened at the
peak compared to the Paczy\'nski curve. 

\MM\, allows the user to make a few choices with regards to the finite source
calculations.  
The limb-darkening coefficient can be provided following the
$u$ convention\footnote{Note that this $u$ is distinct from the $u$ used in Section~\ref{sec:calc1}.  
Unfortunately, common conventions use the same variable name for these two different purposes.} 
(normalized by the central intensity) or the $\Gamma$ convention 
\citep[normalized by the total flux;][]{an02}.  
When fitting the model, the parameter $t_{\star} \equiv \rho t_{\rm E}$ may be 
a better choice than $\rho$, as $t_{\star}$ shows less correlation with other 
light-curve parameters \citep[e.g.][]{yee12}.  The value of $t_{\star}$ can be estimated as the half-width 
of the rounded part of the light curve, e.g., 
$2t_\star \sim 0.1~{\rm d}$ for the example in Figure~\ref{fig:1}.

\subsection{Microlensing parallax} % #######################################
The parallax effect manifests in microlensing as a deflection of 
the trajectory of the lens relative to the source.  
The microlensing parallax $\boldsymbol{\pi_{\rm E}}$ is a 2D vector with magnitude 
equal to the relative lens-source parallax divided by $\theta_{\rm E}$.  
The direction of $\boldsymbol{\pi_{\rm E}}$ is the same as the relative lens-source 
proper motion.  

If $\theta_{\rm E}$ is 
measured \citep[e.g., by measuring finite source effects, see][]{yoo04b}, 
the measurement of the microlensing parallax $\boldsymbol{\pi_{\rm E}}$ allows 
a direct measurement of the lens mass \citep{gould00b}:
$$
M = \frac{\theta_{\rm E}}{\kappa\pi_{\rm E}},
$$
where $\kappa=4G/(c^2{\rm AU})=8.14~{\rm mas~M_{\odot}^{-1}}$ 
(${\rm mas}$ is short for milliarcsecond).  
Microlensing parallax also allows constraining 
the lens distance \citep{gould00b}: 
$$
\frac{1}{D_l} = \pi_{\rm E}\theta_{\rm E} + \frac{1}{D_s}.
$$
The two above equations allow translating parameters from the microlensing model 
(like mass ratio or projected separation of lens components) 
to physical properties of 
the lens system (masses of individual objects in ${\rm M_{\odot}}$ and 
projected separation in ${\rm AU}$).  Hence, measuring, or at least constraining,  
the microlensing parallax plays an important role in characterizing the lens system.  

Below we briefly discuss the microlensing parallax when it manifests as annual 
and satellite effects.  In Figure~\ref{fig:2} we show examples of magnification 
curves affected by both types of the microlensing parallax.  
Note that parallax affects the lens-source relative trajectory and hence 
can be included in both single-lens and double-lens models.  
Some of the model fitting algorithms require $\chi^2$ gradient calculation. 
\MM\, allows calculation of $\chi^2$ gradient for point-source point-lens 
models including parallax models.

\subsubsection{Annual parallax} % ######################################################

The Paczy\'nski curve assumes that the relative lens-source proper motion, 
as seen by the observer, is rectilinear as defined by Equations~\ref{eq:u2} 
and \ref{eq:tau}.  For many microlensing events observed towards 
the Galactic bulge, the timescale $t_{\rm E}$ is shorter than $30~{\rm days}$ 
\citep{wyrzykowski15}.  
During this time, 
Earth's orbital motion can be well-enough approximated as rectilinear.  
For longer events or events with particularly good photometric coverage, 
we may see the effect of Earth's acceleration on the source-lens relative 
trajectory and, hence, in the light curve \citep{an02,gould04b}.  
The calculation of the impact of the annual microlensing parallax 
on the source trajectory requires calculating 
the projection of the Earth's velocity onto the
plane of the sky for a specified reference time for parameters 
$t_{0,{\rm par}}$ \citep{skowron11}.  For point lenses the best choice is
$t_{0,{\rm par}} \approx t_{0}$.  For calculation of relative Earth-Sun 
positions we use high-accuracy ephemerides 
included in {\tt Astropy} package \citep{astropy13,astropy18}.

\subsubsection{Satellite parallax} % ######################################################

Another method to measure the microlensing parallax is to observe 
the same event from at least two well-separated locations 
\citep{refsdal66,gould94c}.  
Typical scales of the bulge microlensing translate to the optimal 
separation of the satellite projected towards the bulge of about 
$1~{\rm AU}$.  Both the \emph{Spitzer} \citep{zhu17b} and 
\emph{K2} \citep{henderson16} satellite missions meet this criterion 
and have observed many microlensing 
events in recent years.  
Calculating the satellite parallax
effect requires calculating the projection of the Earth-satellite
separation on the plane of the sky.  
For extracting accurate satellite positions 
we use JPL Horizons e-mail system 
\citep{giorgini96}\footnote{\url{https://ssd.jpl.nasa.gov/?horizons}} 
and give detailed instructions how this system should be accessed (file {\tt documents/Horizons\_manual.md}).  

\section{Binary lenses} % ######################################################
\label{sec:bin}

A binary lens consists of two components that are often parameterized by their 
mass ratio $q \equiv m_2/m_1$ and their projected separation $s$ (defined as 
a fraction of the Einstein ring radius of the combined mass).  
The magnification calculations for a
binary lens are implemented in the {\tt BinaryLens} class.

\subsection{Lens equation} % ######################################################

The addition of a second mass to the lensing system significantly changes 
the microlensing calculations.  For point lenses discussed in 
Section~\ref{sec:calc}, we treated source and image positions as 
one-dimensional thanks to the symmetry of the problem.  The symmetry is broken 
by the second lens and we have to use two-dimensional coordinates.  
It is most convenient to use complex numbers to represent the coordinates 
\citep{witt90}.  Let $\zeta$, $z$, $z_1$, and $z_2$ be 
the source position, image position, and positions of the two lenses, 
respectively.  The lens equation is then:
$$\zeta = z + \frac{m_1}{\bar{z_1}-\bar{z}} + \frac{m_2}{\bar{z_2}-\bar{z}}.$$
The above equation can be converted 
into a fifth-order complex polynomial and then solved numerically, 
though one has to check if all five solutions of the polynomial are also 
solutions of the lens equation, of which there are either three or five.  
\MM\, uses polynomial coefficients given by \citet{wittmao95} 
and the polynomial is solved using the \citet{bozza18} implementation of the \citet{skowron12} root solver.
The magnification $A$ is 
\begin{equation}
A = \sum\frac{1}{\left|\det J\right|},
\end{equation}
where summation is done over all images, and $J$ is the Jacobian of 
the lens equation.  The example magnification curve for a binary-lens event 
with planetary mass ratio is shown in Figure~\ref{fig:2} 
by the blue solid curve.

The source positions corresponding to $\det J=0$ define a caustic curve, 
on which the point source magnification would be infinite.  
Because the magnification diverges near a caustic, the
physical size of the source and its limb-darkening significantly
affect the total observed magnification within a few source radii
of a caustic.  As noted in Section~\ref{sec:calc2}, for a point lens the
magnification $\rightarrow \infty$ at a single point $u = 0$.  
Thus, for a point lens the caustic is just a point (same as the
lens position), but for binary lenses the caustic curve expands to a
series of edges that describe one, two, or three enclosed regions \citep{erdl93}.  
Hence, the chances of detecting the finite source effect are much higher for 
binary lenses.

\subsection{Finite Source}

The simplest approach to the finite-source binary-lens magnification 
calculations is a direct two-dimensional integration but this 
approach is computationally very inefficient. Far from the caustic, 
it is sufficient to approximate the source as a single point, and 
this is the fastest method for calculating the magnification. Below  
we discuss various approaches to calculating 
the finite-source binary-lens magnification 
in order of increasing accuracy (near a caustic) 
and correspondingly, increasing computation time.  

\subsubsection{Taylor expansion}

After the point source approximation, the next fastest approach is to
calculate the magnification using the quad\-rupole or hexadecapole
approximation of the Taylor expansion of the lensing potential 
\citep{gould08,pejcha09}.  Only nine and thirteen point-source 
lens equation computations are required in the quad\-rupole and hexadecapole approximation, respectively. 
The approximation is generally valid and useful for source positions 
that are between a few and tens of source radii from the caustic. 
As described in \citet{gould08}, the exact region of validity must 
be determined individually for each light curve.  

\subsubsection{Contour integration}

The 2D area of an image of constant surface-brightness source can be
calculated by integrating over the 1D image contour thanks to Green's
theorem \citep{gould97}.  It is faster to integrate over one dimension
compared to two dimensions.  Hence, the contour integration is an
important approach to finite-source binary-lens calculations.

There are several implementation problems of contour integration that
have been solved over the years. {\tt Mulens\-Model} %\MM\, 
implements two of
the available contour integration methods.  
The first method is Adaptive Contouring by \citet{dominik07} 
that uses adaptive grid size to find contours of all images.  
The second method is {\tt VBBL} by \citet{bozza10} and \citet{bozza18} 
that improves the accuracy of integration and
controls the residuals, which allows optimal sampling of
the source. 
In the example cases tested by \citep{bozza10}, the number of lens 
equation solutions was typically a few times larger than the resulting 
magnification for uniform source. The limb darkening effect required 
a few times more lens equation solutions. 
\citet{bozza18} presented a method to 
decide when to use faster point source calculations instead of 
finite source calculations. \MM\, does not use this method at this time, but may add it as an option during future development.
Both Adaptive Contouring and VBBL were implemented by their 
authors in C or C++ and \MM\, calls these codes.

\section{Performance} % ##############################################################################
\label{sec:perf}

To test the code  performance, we benchmark the $\chi^2$ calculation function for 
point-source point-lens event with a single dataset at a time.  The datasets 
had $10^2$, $10^3$, and $10^4$ datapoints. The largest dataset is on the order 
of largest datasets that will be analyzed in foreseeable future. 
We tested both rectilinear motion 
and annual parallax events. To make the benchmark meaningful, we wrote 
short $\chi^2$ function for event without annual parallax using {\tt numpy} package. 
The {\tt numpy}-based function implements Paczy{\'n}ski curve, 
performs linear regression to derive source and blending flux, 
calculates predicted flux values, and evaluates $\chi^2$. 
The goal should be that additional calculations that are performed in the high-level package
\MM\, add as small overhead as possible.

We run the benchmark on a desktop computer with a single socket, 10 core modern processor 
(Intel Xeon E5-2630 v4 2.20GHz). The testing environment was Linux Centos distribution 
with python 3.6.5 (GCC 7.2.0), {\tt numpy} 1.14.3, {\tt scipy} 1.1.0, and {\tt astropy} 3.0.2. 
The testing was performed using 
{\tt perf}\footnote{\url{https://github.com/vstinner/perf}} 
benchmarking package version 1.5.1. 
We run \MM\, twice.  First, the $\chi^2$ contributions of all 
the points are added using accurate floating point sum of values 
(i.e., {\tt math.fsum}), which is the default setting.  Second, the $\chi^2$ contributions
are added using {\tt numpy.sum} function which can be less accurate.
For comparison with \MM\, version 1.4.0 presented here, we also tested {\tt pyLIMA} version 1.0.0. % XXX
We note that {\tt pyLIMA} does not specify conventions for model parameters directly, 
and we had to change signs of $\boldsymbol{\pi_{\rm E}}$ components to obtain results 
consistent with the \citet{skowron11} convention, i.e., 
as if the components are in S and W directions, not N and E. 
The codes used to run the benchmark and raw test results are included in \MM\, distribution 
(see {\tt examples/run\_time\_tests/}). 

\begin{figure}
\includegraphics[width=0.455\textwidth]{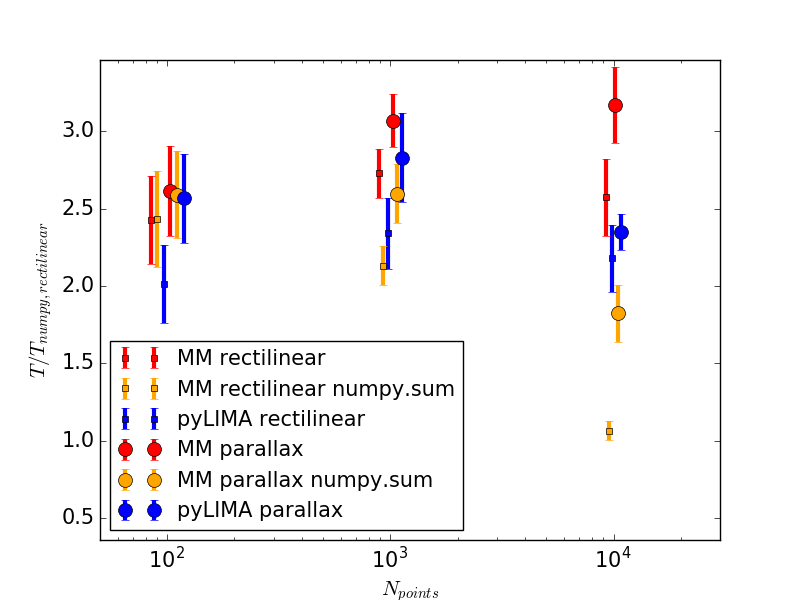}
\caption{Benchmark comparing execution times for \MM\, (red and orange points) 
and {\tt pyLIMA} (blue points). The run time is given relative to run time of 
the simple $\chi^2$ calculation performed in {\tt numpy}. 
Small squares present models with rectilinear motion, 
while large circles indicate annual parallax models.
For \MM\, we show both default calculation (red points) as well as 
faster calculation that uses {\tt numpy.sum} instead of {\tt math.fsum} (orange points).
X axis coordinates are slightly shifted for better visibility.
\label{fig:perf}}
\end{figure}

Figure~\ref{fig:perf} shows ratio of the run time relative to the run time of 
rectilinear model $\chi^2$ calculation with {\tt numpy} and ideally should be close to 1. 
The run times with {\tt numpy} are $0.13~{\rm ms}$, $0.18~{\rm ms}$, and $0.71~{\rm ms}$
for $10^2$, $10^3$, and $10^4$ datapoints, respectively.
In all cases, the ratios are below 3.5. 
For $10^2$ and $10^3$ points, 
the differences between \MM\, and {\tt pyLIMA} 
are comparable to standard deviations. For $10^4$ \MM\, calculation with 
{\tt numpy.sum} is clearly the fastest. 

\section{Additional features} % ######################################################
\label{sec:add}

The main goal of \MM\, is to calculate magnification curves that are
used for model fitting.  The code also provides a number of additional
convenience functions. These functions fall outside the main scope
of the code, but the user may find them useful. Thus, we describe
them here.

\subsection{Plotting}

\begin{figure}
\includegraphics[width=.95\columnwidth]{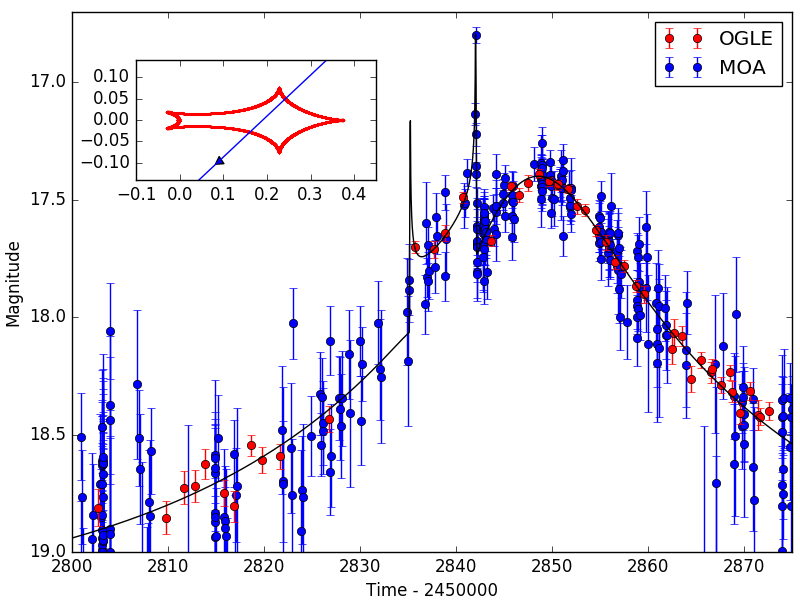}
\caption{Data for OGLE-2003-BLG-235/MOA-2003-BLG-53 shown with a model 
generated with \MM. The model is based on \citet{bond04}, but the exact 
parameters were adjusted to better approximate the data. Only 
the flux parameters were fit by \MM.
Photometry was acquired from NASA Exoplanet Archive.  
The inset shows the source trajectory (blue) relative to caustic (red). 
\label{fig:3}}
\end{figure}

\MM\, offers several built-in plotting functions to
facilitate the visualization of models, data, and the model
residuals.  In addition to the magnification curves shown in
Figures~\ref{fig:1} and \ref{fig:2}, \MM\, makes it easy for the user to plot data
with a given model.  For each dataset the optimum values of $F_S$ and $F_B$
are found and the brightness measurements are scaled to
the same magnitude system (Equation~\ref{eq:flux}).
Figure~\ref{fig:3} shows the OGLE and MOA
data\footnote{These data were downloaded from the NASA Exoplanet Archive \url{https://exoplanetarchive.ipac.caltech.edu/}}
for OGLE-2003-BLG-235/MOA-2003-BLG-53 \citep{bond04} and the model light curve.
There are also built-in functions for plotting
the caustics and the trajectory of the source relative to the lens.

\subsection{Reparameterization}

There are multiple sets of parameters that can be used for model fitting
and the optimal choice depends on the specific event being considered.
In addition to $\rho \leftrightarrow t_{\star}$ reparameterization described
in Section~\ref{sec:calc2},
\MM\, allows use of the effective timescale ($t_{\rm eff} \equiv u_0t_{\rm E}$)
instead of either $u_0$ or $t_{\rm E}$.

\subsection{Hypothetical Systems}

The {\tt mulensobjects} submodule allows the user to easily calculate
microlensing quantities by defining a physical lens-source
system.  For example, the user may specify a lens mass, distances to
the source and lens, and the relative lens-source proper motion and
then retrieve $t_{\rm E}$ or various projections of the Einstein
radius.

\section{Future development} % ######################################################
\label{sec:fut}

While version 1.4.0 of \MM\, allows a wide variety of events to % XXX
be analyzed, we can envision many avenues for updates and expansions
to the code, some of which we briefly outlined below.

\subsection{Additional magnification calculation methods}

There are several other methods for calculating the source
magnification when finite source effects are significant.  These may be
more accurate or more efficient than those currently implemented,
depending on the exact case.  For example, for point lenses in which
$\rho \gtrsim 0.1$, the \citet{lee09} method should be used.  In the
case of binary lenses, there are other contour integration algorithms
\citep[e.g. ][]{dong09b}. 
In addition to contour integration, the finite-source binary-lens magnification 
can be calculated using the inverse-ray shooting method \citep{kayser86}.  
The details of the inverse-ray shooting method have been improved over the years 
\citep[e.g. ][]{vermaak00,dong06,bennett10b}. 
Recently, \citet{cassan17} proposed a method to increase 
efficiency of Taylor expansion calculations. 
Implementing some or all of these methods could be a direction for 
future development.  

\subsection{Additional parameterizations}

\MM\, is built using the center-of-mass coordinate
system.  However, there are a number of other possible
parameterizations of the event that can make fitting easier and
more efficient.  For example, \citet{cassan08} proposed a binary lens
parameterization that uses the epochs of the caustic crossings, which are
well measured, and two coordinates along the caustic 
as fitting parameters in place of, 
e.g., $t_0$, $u_0$, $t_{\rm E}$, and $\alpha$.  
\citet{penny14} presented a way to optimize planetary events simulations. 
The properties of caustics in planetary mass ratio regime were studied 
by \citet{chung05} and \citet{han06}.  The parameters derived by them 
can be useful in fitting particular events 
and in some cases one can estimate event properties based on 
a simple light curve inspection \citep{gould92,poleski14a}.
Even for point-lens events one may want to use different sets
of microlensing parameters to speed-up the calculations \citep{yee12}.
Re-parameterization not only makes exploration of the parameter 
space easier, but in doing so improves the probability that all
alternative, degenerate models have been found and considered.  Since
the user defines the minimization algorithm, the user can define a
likelihood function that converts an arbitrary parameter set into the
center-of-mass coordinate system used by \MM\, and then
calculate and return the $\chi^2$. However, a path for future
development would be to create built-in functions to convert between
various parameterizations.

\subsection{Higher-order Effects and Triple Lenses}

There are additional, higher-order effects that we plan to
implement in the near future: binary
sources (two luminous sources), the xallarap effect
\citep{griest92,dominik98a}, and two-parameter limb-darkening.  
We note that currently there is no capability in \MM\, for
astrometric microlensing calculations \citep{belokurov02,sahu17}
but it can be added in future.  

Another obvious path for future development is adding triple-lens
models. The serious limitation in implementing the finite-source
triple-lens models is the lack of deep understanding of triple-lens
caustic structures \citep[see, e.g.,][]{danek15b,luhn16}.
For numerical contour integration methods, this leads to problems with 
correctly matching up the points along the contours. The problem is 
particularly severe near 
swallowtail and butterfly morphologies.  
Also the model degeneracies are more severe \citep{song14}.  
Additionally, in some cases, the solutions of 10-th order polynomial can be 
numerically unstable \citep{han02e,bennett10b}.

\subsection{\emph{WFIRST} Data Analysis Challenges}

A series of Data Analysis 
Challenges\footnote{\url{http://microlensing-source.org/data-challenge/}} 
in advance of \emph{WFIRST} are currently
underway, and we hope that {\tt Mu\-lens\-Model} % \MM\, 
will serve as a useful
tool for those challenges and for development of the \emph{WFIRST}
analysis pipeline. The immediate, future development of \MM\, 
will likely be driven by ensuring that \MM\,
can be used to address the problems posed by the Data Analysis Challenges. 
Thus, most likely, the next round of development will focus on implementing
orbital motion, additional model parameterizations, and triple
lenses. We also anticipate that the Data Analysis Challenges will reveal new
pathways for future development on \MM. 

We welcome community feedback on the current status of the code,
requests of features to be added, and help in developing the code.

\section*{Acknowledgements} % ######################################################

This work was supported by NASA contract NNG16PJ32C.  
We thank \emph{WFIRST} Microlensing Science Investigation Team for consultation.  
The SciCoder Workshop organized by Demitri Muna is acknowledged.  
This research has made use of the NASA Exoplanet Archive, which is
operated by the California Institute of Technology, under contract
with the National Aeronautics and Space Administration under the
Exoplanet Exploration Program.

\appendix
\section{Source codes}
\label{sec:appA}

Below we present the source codes used to prepare 
the Figures~\ref{fig:1}, \ref{fig:2}, and \ref{fig:3}.  
The figures include a few higher order effects 
and subpanels presenting caustics, 
yet the codes are succinct.

\begin{lstlisting}[language=Python,frame=single,float=*,caption={Code used to prepare Figure~\ref{fig:1}}]
"""
Create Figure 1.

Example point-source magnification curves.

"""
from matplotlib import pyplot

from MulensModel import Model


# Define model parameters.
t_0 = 2460000
params_ps = {'t_0': t_0, 'u_0': 0.0008, 't_E': 30.}
t_star = 0.051  # Day is default unit for t_E and t_star.
gamma = 0.4  # This is limb darkening coefficient.
params_fs = {**params_ps, 't_star': t_star}

# Set models settings for:
model_ps = Model(params_ps)  # point source,
model_fs = Model(params_fs)  # finite source,
model_fs_ld = Model(params_fs)  # and finite source with limb darkening.

# Specify which finite source methods are used and when:
t_1 = t_0 - 3.5 * t_star
t_2 = t_0 + 3.5 * t_star
model_fs.set_magnification_methods([t_1, 'finite_source_uniform_Gould94', t_2])
model_fs_ld.set_magnification_methods([t_1, 'finite_source_LD_Yoo04', t_2])

# Plot the magnification curves.
plot_kwargs = {'t_start': t_0-5.5*t_star, 't_stop': t_0+5.5*t_star,
               'subtract_2460000': True, 'lw': 2.}
model_ps.plot_magnification(label='point source', **plot_kwargs)
model_fs.plot_magnification(label='finite source', **plot_kwargs)
model_fs_ld.plot_magnification(
    gamma=gamma, label='finite source LD', **plot_kwargs)

pyplot.legend(loc='best')
pyplot.savefig('figure_1.png')
\end{lstlisting}

\begin{lstlisting}[language=Python,frame=single,float=*,caption={Code used to prepare Figure~\ref{fig:2}}]
"""
Create Figure 2.

Example magnification curves.

"""
from matplotlib import pyplot
import os

from MulensModel import Model, SatelliteSkyCoord, MODULE_PATH


# Define model parameters.
params = {'t_0': 2456900, 'u_0': 0.2, 't_E': 50.}
params_pi_E = {'pi_E_N': 0.35, 'pi_E_E': 0.5}
params_planet = {'rho': 0.002, 's': 1.5, 'q': 0.001, 'alpha': 348.1}
ra_dec = '18:00:00.00 -28:30:00.0'

# Set models and satellite settings.
model_pspl = Model(params)
model_planet = Model({**params, **params_planet})

# Calculate finite source magnification using VBBL method for this
# range of dates:
model_planet.set_magnification_methods([2456937, 'VBBL', 2456945])

# Parallax settings:
model_parallax = Model({**params, **params_pi_E}, coords=ra_dec)
model_parallax.parallax(earth_orbital=True, satellite=True)
satellite = SatelliteSkyCoord(
    os.path.join(
        MODULE_PATH, 'data/ephemeris_files', 'Spitzer_ephemeris_01.dat'))
# This file gives the Spitzer ephemeris and is part of MulensModel package.

# Plot the magnification curves.
plot_kwargs = {'subtract_2450000': True, 'lw': 2.}
pyplot.figure(figsize=(8,8))
pyplot.axes([0.1, 0.43, 0.85, 0.55])
model_planet.plot_magnification(label='planetary', **plot_kwargs)
model_parallax.plot_magnification(
    label='annual parallax', linestyle='-.', **plot_kwargs)
model_pspl.plot_magnification(label='PSPL', linestyle='--', **plot_kwargs)
model_parallax.plot_magnification(
    label='satellite parallax', satellite_skycoord=satellite, **plot_kwargs)
pyplot.legend(loc='best')

pyplot.axes([0.1, 0.07, 0.85, 0.25]) # Lower panel starts here.
model_planet.plot_trajectory(caustics=True)
pyplot.xlim(-1.52, 1.61)
pyplot.xlabel(r"$\theta_x$")
pyplot.ylabel(r"$\theta_y$")
pyplot.savefig('figure_2.png')
\end{lstlisting}

\begin{lstlisting}[language=Python,frame=single,float=*,caption={Code used to prepare Figure~\ref{fig:3}}]
"""
Creates Figure 4.

This example shows OGLE-2003-BLG-235/MOA-2003-BLG-53,
the first microlensing planet.

"""
from matplotlib import pyplot
import os

from MulensModel import MulensData, Model, Event, MODULE_PATH


# Import data
data_dir = os.path.join(MODULE_PATH, 'data', 'photometry_files', 'OB03235')

OGLE_data = MulensData(
    file_name=os.path.join(data_dir, 'OB03235_OGLE.tbl.txt'),
    comments=['\\', '|'])
MOA_data = MulensData(
    file_name=os.path.join(data_dir, 'OB03235_MOA.tbl.txt'),
    comments=['\\', '|'], phot_fmt='flux')

# Define a model with a 2-body lens (these parameters slightly differ
# from Bond et al. 2004):
model_1S2L = Model({'t_0': 2452848.06, 'u_0': 0.1317, 't_E': 61.5,
                    'rho': 0.00096, 'q': 0.0039, 's': 1.120, 'alpha': 223.72})

# Since rho is set, define a time range and method to apply finite
# source effects:
model_1S2L.set_magnification_methods([2452833., 'VBBL', 2452845.])

# Combine the data and model into an Event:
my_event = Event(datasets=[OGLE_data, MOA_data], model=model_1S2L)

# Make the plot:
t_range = [2452800., 2452875.]
pyplot.axes([0.09, 0.08, 0.9, 0.9])
my_event.plot_data(
    subtract_2450000=True, label_list=['OGLE', 'MOA'],
    color_list=['red', 'blue'], zorder_list=[2, 1], s=6)
my_event.plot_model(
    subtract_2450000=True, t_range=t_range, n_epochs=4000, color='black')

pyplot.legend(loc='best')
pyplot.xlim(t_range[0]-2450000., t_range[1]-2450000.)
pyplot.ylim(19.0, 16.7)

pyplot.axes([0.17, 0.7, 0.3, 0.2]) # Figure inset stars here.
model_1S2L.plot_trajectory(caustics=True)
pyplot.xlim(-0.1, 0.45)
pyplot.ylim(-0.14, 0.14)
pyplot.savefig('figure_4.png')
\end{lstlisting}

\section{Public microlensing datasets}
\label{sec:appB}

\MM\, is meant to be used for analysis of real microlensing events
but also exploring novel solutions to the computational challenges
faced by microlensing. Conducting such an exploration requires
access to photometric data of real microlensing events. There now
exist several public data sets, which might be analyzed or used to
study these problems. Because a list of these data sets has not been previously compiled, we review here the public time-series 
photometry of microlensing events starting from newest datasets. 

The KMTNet survey has operated telescopes on three continents since 2015 
\citep{kim16}.  The photometry of events from 2015 commissioning season 
and 2016 data for \emph{K2} Campaign 9 footprint are public 
\citep{kim18a,kim18b} (altogether 841 clear events and 266 classified as possible) 
and future datasets will also be publicly released.  
In 2015 and 2016 the microlensing survey was conducted using UKIRT telescope 
\citep{shvartzvald17a} and all $18\times10^6$ aperture photometry light curves were 
released\footnote{\url{https://exoplanetarchive.ipac.caltech.edu/docs/UKIRTMission.html}}. 
Similarly, aperture photometry for all sources observed between 2010 and 2015 
by the VVV survey \citep{minniti10} is also 
public\footnote{\url{http://archive.eso.org/cms/eso-archive-news/New_Data_Release_of_VVV_Photometric_Catalogues_via_the_ESO_Science_Archive_Facility.html}}.

The public time-series images are available for events observed by 
\emph{Spitzer} and \emph{K2} satellite missions.  \emph{Spitzer} has conducted 
microlensing campaigns since 2014, while \emph{K2} conducted the first space-based 
microlensing survey in its Campaign 9 \citep{henderson16} 
with additional targets observed in Campaign 11.  Extracting photometry from both 
\emph{Spitzer} and \emph{K2} images of Galactic bulge requires specialized techniques  
\citep{calchinovati15b,zhu17a}\footnote{See also \url{https://github.com/CPM-project/MCPM}}. 

The sample of 3718 OGLE-III survey \citep{udalski08red} 
events used for Galactic bulge structure study was 
presented by \citet{wyrzykowski15}.  These events are primarily point-source 
point-lens events, but parallax events and some binary events are present 
in this sample. 
The MOA-II survey \citep{sako08} sample consists of 474 events \citep{sumi13} 
and some of these are common with \citet{wyrzykowski15} sample. 
Similarly, \citet{sumi06} 
released 122 light-curves from the OGLE-II survey.  
A sample of 214 OGLE-II light-curves was published by \citet{udalski00a}.  
Additional OGLE-II microlensing events are present in the \citet{wozniak02} 
catalog of $2\times10^5$ variable sources. 
Also \citet{thomas05} released 564 light-curves from the MACHO project bulge study. 

The photometry of events that were published as planetary microlensing 
can be accessed via the NASA Exoplanet Archive. 

%% Authors are advised to submit their bibtex database files. They are
%% requested to list a bibtex style file in the manuscript if they do
%% not want to use model2-names.bst.

\bibliographystyle{model2-names}
\bibliography{paper}

\end{document}